\documentclass[submission,copyright,creativecommons]{eptcs}
\providecommand{\event}{Wivace 2013}
\usepackage[utf8]{inputenc}
\usepackage{graphicx}
\usepackage{url}
\usepackage{breakurl}

\title{A Model of Colonic Crypts using SBML Spatial}
\author{Daniele Ramazzotti
\institute{Dipartimento di Informatica, Sistemistica e Comunicazione\\
Università di Milano-Bicocca\\
Milan, Italy}
\email{daniele.ramazzotti@disco.unimib.it}
\and
Carlo Maj
\institute{Dipartimento di Informatica, Sistemistica e Comunicazione\\
Università di Milano-Bicocca\\
Milan, Italy}
\email{carlo.maj@disco.unimib.it}
\and
Marco Antoniotti
\institute{Dipartimento di Informatica, Sistemistica e Comunicazione\\
Università di Milano-Bicocca\\
Milan, Italy}
\email{marco.antoniotti@disco.unimib.it}
}
\def\titlerunning{A Model of Colonic Crypts using SBML Spatial}
\def\authorrunning{D. Ramazzotti, C. Maj \& M. Antoniotti}

\begin{document}
\maketitle

\begin{figure}[htbp]
\begin{center}
\includegraphics[width=0.6\textwidth]{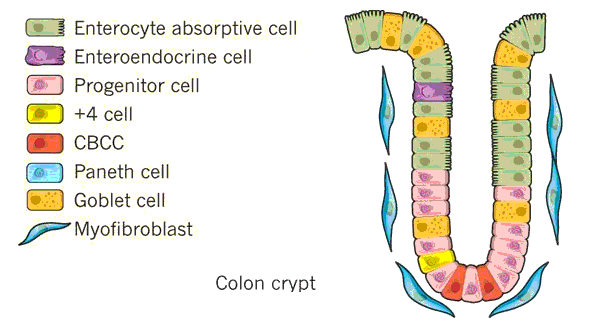}
\end{center}
\caption{Schematic representation of a colonic crypt from\cite{Medema}. In the figure, 
on the left the cellular types present in the crypt are listed, while on the right the shape 
of the crypt is represented.}
\label{colonic_crypt}
\end{figure}

Colonic crypts are invaginations of the connective tissue of human intestine and 
are supposed to be the site where mutations affecting the stem cells can occur 
leading to the emergence and progression of Colorectal Cancer (CRC)\cite{De_Matteis}. 
See figure \ref{colonic_crypt} for a schematic representation of a colonic crypt from\cite{Medema}.

Different models aimed at describing the colonic crypt behavior have
been defined over the years and can be divided in two groups: 
\emph{in-lattice models}, \cite{Chaste}, \cite{CellSys-Drasdo-Hoeme}, 
and \emph{off-lattice}, \cite{CompuCell3D}. See \cite{De_Matteis} 
and further reference therein for a recent review on the subject. 

In this work we propose a computational model that implements the most recent SBML 
format \cite{Core} and, in particular, the SBML \emph{Spatial Processes} package. 
SBML is currently the most used standard for representing computational models in 
systems biology. It is open and has widespread software support and a community 
of users and developers. 

The \emph{Spatial Processes}  package introduces a  new tag, named \emph{geometry}, 
which enables an explicit definition of a spatial environment for the simulation. The 
possibility of an explicit representation of spatial dynamics increases the representation 
power of SBML. The SBML \emph{Spatial Processes} package has been proposed in 2010 and 
is currently still under development. At March 2013 the latest version of this 
package is the 0.81 release of July 2012\cite{Combine}. In this work, we refer 
to this version of SBML \emph{Spatial Processes}.\footnote{In the \emph{geometry} tag, five 
new SBML subtags are permitted:
\begin{itemize}
 \item \textit{ListOfCoordinateCompartments}: in this sub-tag the spatial 
 frame is defined. Different types of reference frames are permitted: 
 in our model it is a 3-dimensional Cartesian System where the x-axis represent 
 the width, the y-axis the height and the z-axis the depth of the geometrical shapes.
 \item \textit{ListOfDomainTypes}: in this sub-tag, homogeneous spatial zones present 
 in the system should be defined. Each spatial zone is intended as being 
 anatomically and physiologically similar and the domain types defined in this 
 tag can refer to one or multiple concrete domains (defined in the next tag). For instance, 
 in our case the cell domain types have been defined here, while the single concrete cells 
 have been defined in the next tag.
 \item \textit{ListOfDomains}: the domains represent contiguous regions identified by 
 the same domain type. For each domain a position in the reference frame defined before 
 is assigned. The domains defined here should match the initial condition of the 
 dynamic model.
 \item \textit{ListOfAdjacentDomains}: adjacent domain types can be defined here two 
 by two. In our case we have a domain type for each cell position in the crypt hence each 
 domain type has multiple adjacent domains.
  \item \textit{ListOfGeometryDefinitions}: here is defined the geometrical structure of 
  each domain type. This is an abstract structure to be assigned to the real domains 
  linked through the domain types definition. SBML \emph{Spatial Processes} offers 
  four possible ways to define the geometry: in our case the \emph{AnalyticalGeometry} 
  tag has been adopted\vspace{-3ex}.
\end{itemize}}\\

Our goal is to model the dynamics and the spatial evolution of the tissue by taking 
into account both the cellular differentiation and cellular migration processes in 
specific crypt locations. Our model consists of two main parts: the dynamic component, which
models the cellular differentiation process, and the spatial component, which
models the positioning and the movement of cells.\\
 
The basic dynamic part of the model describes 8 cellular types and 12 cellular transformations. 
The representation with SBML core has been done by describing the cellular types as species and 
the cellular transformation as reactions. In addition to these 8 species, an \emph{empty} 
cell is considered in order to represent the empty space in the colonic crypt.

More in details, the 8 cellular types are Stem cell, Paneth cell,  Ta1 (mutated cells 
of type 1), Ta2a (mutated cells of type 2-a), Ta2b (mutated cells of type 2-b), Goblet 
cell, Enteroendocrine cell and Enterocyte absorptive cell. The choice of these cellular 
types has been done to represent the cellular differentiation processes that gradually 
transforms the stem cell progeny into four different fully differentiated cellular types 
(paneth, goblet, enteroendocrine and enterocyte). Three types of partially 
differentiated cellular types are present (ta1, ta2a, ta2b) in our model.

\begin{figure}[htbp]
\begin{center}
\includegraphics[width=0.5\textwidth]{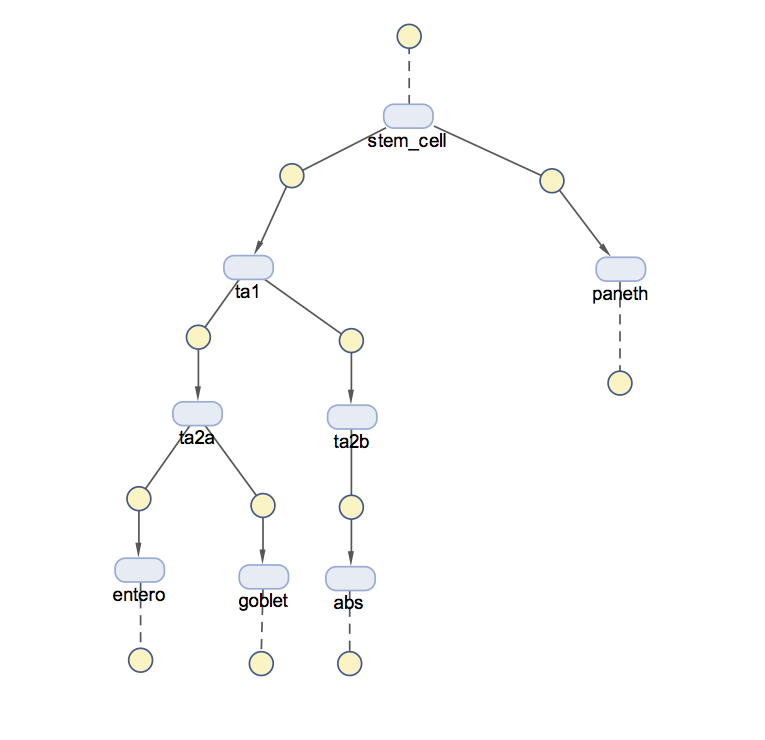}
\end{center}
\caption{Model Network: the resulting network generated by the model. It involves 
the differentiated cell types (paneth, goblet, enteroendocrine and enterocyte) and 
three partially differentiated cells (ta1, ta2a, ta2b).}
\label{network}
\end{figure}

We have defined twelve reactions: seven reactions to represent the cellular differentiations, 
one reaction regarding the duplication of the stem cells and four reactions to represent 
the degradation of  the final differentiated cell types (paneth, goblet, enteroendocrine and 
enterocyte). The duplication and the degradation reactions are present in order to assure 
the attainment of a steady condition. More in detail, the duplication of the stem cells assures 
that the system does not cease to work as a consequence of the depletion of stem cells, 
while the degradation of the differentiated cell prevents the unlimited formation of the 
terminal node of the network. In figure \ref{network} the resulting dynamic network is shown.

Once the spatial system has been defined, it is linked to and guided by the 
dynamic part of the system to reflect the spatial evolution of the colonic crypts 
(i.e. downward displacement of the Paneth cells and upward movement of the other 
cells once they specify from stem cells).\\

\begin{figure}[htbp]
\begin{center}
\includegraphics[width=0.6\textwidth]{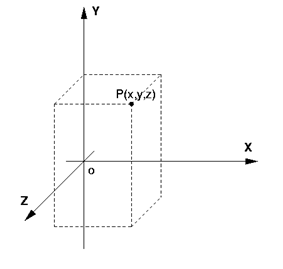}
\end{center}
\caption{In our model the space is discretized in a finite number of cells following an 
in-lattice approach.}
\label{cell}
\end{figure}

The spatial part of the model exploits an in-lattice representation
where the colonic crypt is described with a series of cubic cells
which can be empty or filled with one of the 8 species defined in the
dynamic model. The colonic crypt has been represented as a hollow parallelepiped 
placed in a 3-dimensional xyz Cartesian reference frame. The width and depth of 
the parallelepiped are represented respectively by the x and z axis. The y-axis 
represents the height of the parallelepiped. The spatial dynamics of the
system consists in upward and downward movements, hence the y-values can 
be within a certain range defined by the dynamic itself. See figure \ref{cell}.

\begin{figure}[htbp]
\begin{center}
\includegraphics[width=0.4\textwidth]{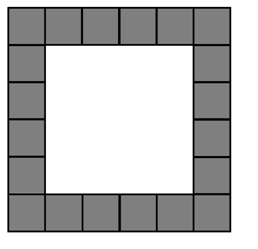}
\end{center}
\caption{A Section of the Colonic Crypt from above, showing the positioning of the cells and 
that the interior area is empty.}
\label{section}
\end{figure}

Figure \ref{section} shows a section of the crypt as represented in our model with a 
view from above.

Finally, three special layers have been defined in the crypt. The first one is located in the 
middle-lower part of the crypt and here new stem cells are created in order to feed the system 
and avoid the possibility of completly consume all the stem cells. Then two more special layers 
are defined at the lower and upper border of the system in order to consume the cells that 
are leaving the crypt.\\

This work is a first attempt to attach the task of using SBML 
\emph{Spatial Processes} to model colonic crypts. The latest \emph{VCell} 
release\cite{VCELL}, which is the first modeling software implementing 
the extension package, has been adopted to define and import the model.

Currently, \emph{VCell} is under developement and its spatial simulation 
capabilities will be improved in order to allow more powerfull simulations 
and parameter tuning, which will be the next steps of this work.\\
 
Ongoing simulations are aimed at defining the parameter tuning for 
the model parameters such, i.e. the initial conditions and the reaction rates that better 
represents the colonic crypt dynamics and allows the maintenance of a stable dynamic state, 
i.e. homeostasis.

Besides, further simulations will be performed in order to verify the robustness and the 
homeostasis of the system. This will be accomplished by analysing the influence of a variation of 
the initial conditions, as well as perturbations of other key parameters of the model.

\nocite{*}
\bibliographystyle{eptcs}
\bibliography{biblio}

\begin{thebibliography}{1}
\providecommand{\bibitemdeclare}[2]{}
\providecommand{\surnamestart}{}
\providecommand{\surnameend}{}
\providecommand{\urlprefix}{Available at }
\providecommand{\url}[1]{\texttt{#1}}
\providecommand{\href}[2]{\texttt{#2}}
\providecommand{\urlalt}[2]{\href{#1}{#2}}
\providecommand{\doi}[1]{doi:\urlalt{http://dx.doi.org/#1}{#1}}
\providecommand{\bibinfo}[2]{#2}

\bibitemdeclare{misc}{CompuCell3D}
\bibitem{CompuCell3D}
\emph{\bibinfo{title}{CompuCell3D}}.
\newblock \bibinfo{howpublished}{\url{http://www.compucell3d.org/}}.

\bibitemdeclare{misc}{VCELL}
\bibitem{VCELL}
\emph{\bibinfo{title}{Virtual Cell Modeling and Analysis Software}}.
\newblock \bibinfo{howpublished}{\url{http://www.nrcam.uchc.edu/}}.

\bibitemdeclare{article}{De_Matteis}
\bibitem{De_Matteis}
\bibinfo{author}{Giovanni \surnamestart De~Matteis\surnameend},
  \bibinfo{author}{Alex \surnamestart Graudenzi\surnameend} \&
  \bibinfo{author}{Marco \surnamestart Antoniotti\surnameend}
  (\bibinfo{year}{2012}): \emph{\bibinfo{title}{A review of spatial
  computational models for multi-cellular systems, with regard to intestinal
  crypts and colorectal cancer development}}.
\newblock {\sl \bibinfo{journal}{Journal of Mathematical Biology}}, pp.
  \bibinfo{pages}{1--54}, \doi{10.1007/s00285-012-0539-4}.

\bibitemdeclare{manual}{Core}
\bibitem{Core}
\bibinfo{author}{Michael \surnamestart Hucka\surnameend},
  \bibinfo{author}{Frank~T. \surnamestart Bergmann\surnameend},
  \bibinfo{author}{Stefan \surnamestart Hoops\surnameend},
  \bibinfo{author}{Sarah~M. \surnamestart Keating\surnameend},
  \bibinfo{author}{Sven \surnamestart Sahle\surnameend},
  \bibinfo{author}{James~C. \surnamestart Schaff\surnameend},
  \bibinfo{author}{Lucian~P. \surnamestart Smith\surnameend} \&
  \bibinfo{author}{Darren~J. \surnamestart Wilkinson\surnameend}
  (\bibinfo{year}{2010}): \emph{\bibinfo{title}{The Systems Biology Markup
  Language (SBML): Language Specification for Level 3 Version 1 Core}}.

\bibitemdeclare{article}{Chaste}
\bibitem{Chaste}
\bibinfo{author}{Pitt-Francis \surnamestart J\surnameend},
  \bibinfo{author}{Pathmanathan \surnamestart P\surnameend},
  \bibinfo{author}{Bernabeu \surnamestart MO\surnameend},
  \bibinfo{author}{Bordas \surnamestart R\surnameend}, \bibinfo{author}{Cooper
  \surnamestart J\surnameend}, \bibinfo{author}{Fletcher \surnamestart
  AG\surnameend}, \bibinfo{author}{Mirams \surnamestart GR\surnameend},
  \bibinfo{author}{Murray \surnamestart P\surnameend}, \bibinfo{author}{Osborne
  \surnamestart JM\surnameend}, \bibinfo{author}{Walter \surnamestart
  A\surnameend}, \bibinfo{author}{Chapman \surnamestart SJ\surnameend},
  \bibinfo{author}{Garny \surnamestart A\surnameend},
  \bibinfo{author}{\surnamestart van Leeuwen~IM\surnameend},
  \bibinfo{author}{Maini \surnamestart PK\surnameend},
  \bibinfo{author}{RodrÌguez~B \surnamestart andWaters SL\surnameend},
  \bibinfo{author}{Whiteley \surnamestart JP\surnameend},
  \bibinfo{author}{Byrne \surnamestart HM\surnameend} \&
  \bibinfo{author}{Gavaghan \surnamestart DJ\surnameend}:
  \emph{\bibinfo{title}{Chaste: A test-driven approach to software development
  for biological modelling}}.
\newblock {\sl \bibinfo{journal}{Computer Physics Communications}}
  \bibinfo{volume}{180(12)}, p. \bibinfo{pages}{2452–2471},
  \doi{10.1016/j.cpc.2009.07.019}.

\bibitemdeclare{article}{Combine}
\bibitem{Combine}
\bibinfo{author}{N.~\surnamestart Le~Novere\surnameend},
  \bibinfo{author}{M.~\surnamestart Hucka\surnameend},
  \bibinfo{author}{N.~\surnamestart Anwar\surnameend}, \bibinfo{author}{G.D.
  \surnamestart Bader\surnameend}, \bibinfo{author}{E.~\surnamestart
  Demir\surnameend}, \bibinfo{author}{S.~\surnamestart Moodie\surnameend} \&
  \bibinfo{author}{A.~\surnamestart Sorokin\surnameend} (\bibinfo{year}{2011}):
  \emph{\bibinfo{title}{Meeting report from the first meetings of the
  Computational Modeling in Biology Network (COMBINE)}}.
\newblock {\sl \bibinfo{journal}{Standards in genomic sciences}}
  \bibinfo{volume}{5}(\bibinfo{number}{2}), p. \bibinfo{pages}{230}.
\doi{10.4056/sigs.2034671}

\bibitemdeclare{article}{Medema}
\bibitem{Medema}
\bibinfo{author}{J.P. \surnamestart Medema\surnameend} \&
  \bibinfo{author}{L.~\surnamestart Vermeulen\surnameend}
  (\bibinfo{year}{2011}): \emph{\bibinfo{title}{Microenvironmental regulation
  of stem cells in intestinal homeostasis and cancer}}.
\newblock {\sl \bibinfo{journal}{Nature}}
  \bibinfo{volume}{474}(\bibinfo{number}{7351}), pp. \bibinfo{pages}{318--326}.
\doi{10.1038/nature10212}

\bibitemdeclare{misc}{CellSys-Drasdo-Hoeme}
\bibitem{CellSys-Drasdo-Hoeme}
\bibinfo{author}{Harrison \surnamestart NC\surnameend} (\bibinfo{year}{2012}):
  \emph{\bibinfo{title}{BioCellSim 2.0}}.
\newblock
  \bibinfo{howpublished}{\url{http://pcwww.liv.ac.uk/~mf0u4027/biocellsim}}.

\end{thebibliography}

\end{document}